\documentclass[epj]{webofc}
\usepackage[varg]{txfonts}   % Web of Conferences font
\newcommand{\SB}{\mathrm{SB}}
\wocname{EPJ Web of Conferences}
\woctitle{CONF12}
\begin{document}
\selectlanguage{english}
\title{Frontiers of finite temperature lattice QCD}
%
% subtitle (optional, strongly discouraged)
%
%%%\subtitle{Do you have a subtitle?\\ If so, write it here}

\author{Szabolcs Bors\'anyi\inst{1}\fnsep\thanks{\email{borsanyi@uni-wupperta.l.de}} %\and
%        Second author\inst{2} \and
%        Third author\inst{3}
        % etc.
}

\institute{Theoretical Physics, University of Wuppertal, 42119 Wuppertal, Germany
%\and
%           The second here 
%\and
%           The last address here
}

\abstract{%
I review a selection of recent finite temperature lattice results of the past
years.  First I discuss the extension of the equation of state towards high
temperatures and finite densities, then I show recent results on the QCD
topological susceptibility at high temperatures and highlight its relevance for
dark matter search.
}
\maketitle
\section{Introduction}
\label{sec:intro}
Quantum chromodynamics (QCD) is the underlying quantum field theory
of the strong interaction that is probed in heavy ion experiments,
such as the Relativistic Heavy Ion Collider (RHIC) at Brookhaven,
and at the Large Hadron Collider (LHC) at CERN. The relevance of
QCD predictions, however, stretches out to temperatures much beyond
the reach of these experiments, and must be considered in cosmological
scenarios of the Early Universe \cite{diCortona:2015ldu}.

Lattice simulations provide an excellent tool to study equilibrium
QCD at all but the perturbative temperature ranges of interest.
The algorithmic difficulties that prevented the use of the light
physical quark masses have been mostly overcome \cite{Aoki:2012oma}, and the
systematics of the lattice discretization is under control by using 
sufficiently fine lattices and performing a continuum extrapolation,
allowing lattice QCD to reconstruct the hadron spectrum 
\cite{Durr:2008zz,Borsanyi:2014jba}.

Thermodynamics calculations are also possible with physical quark masses with
either staggered \cite{Aoki:2005vt} or domain wall fermions
\cite{Bhattacharya:2014ara}. The transition is a cross-over \cite{Aoki:2006we},
its temperature has been calculated based on the chiral observables,
$T_c\approx 155~\mathrm{MeV}$
\cite{Aoki:2006br,Aoki:2009sc,Borsanyi:2010bp,Bazavov:2011nk}, the
deconfinement aspects did not allow a clear-cut $T_c$ definition, though all
possible definitions are within the peak range of the chiral susceptibility
\cite{Borsanyi:2010bp}.  Recently the single-quark entropy was introduced based
on the Polyakov loop which also peaks near the chiral $T_c$
\cite{Bazavov:2016uvm}.  The equation of state has been determined well beyond
the transition regime, including the temperature derivatives, such as the trace
anomaly and higher derivatives, like the speed of sound and heat capacity
\cite{Borsanyi:2010cj,Borsanyi:2013bia,Bazavov:2014pvz}. In quenched QCD the
temperature range stretches out into the perturbative regime
\cite{Borsanyi:2012ve,Giusti:2016iqr}.

Several experimental as well as lattice papers have been lately
devoted to the fluctuations of conserved charges
\cite{Borsanyi:2015axp,Asakawa:2015ybt}.  In a grand canonical ensemble (with
zero or non-zero chemical potentials) conserved quantum numbers, such as the
baryon number, electric charge or strangeness, fluctuate according to
equilibrium thermodynamic features, that are given by even derivatives of the
free energy with respect to the corresponding chemical potentials. The odd
derivatives vanish due to the C-symmetry of the QCD action. The magnitude of
these fluctuations are distinctive features
of the two phases around the QCD transition \cite{Koch:2012psa}. Precision
experimental studies provide the values of ratios of such fluctuations for
several collision energies as part of the RHIC beam energy scan program
\cite{Adamczyk:2013dal,Adamczyk:2014fia}.  The obtained data can be described
by a grand canonical ensemble assuming a temperature and chemical potential
which are referred to as chemical freeze-out parameters. This is a
generalization of the earlier statistical hadronization model, which was based
on particle yields instead of fluctuations
\cite{Cleymans:2005xv,Andronic:2005yp}.  The fluctuation-based freeze-out
temperatures lie in general below the original yield-based freeze-out curve
\cite{Alba:2014eba}.

These achievements of lattice QCD are mostly limited to zero chemical potential
and to near-transition temperatures. In this work we describe the progress
towards finite densities, relevant for RHIC phenomenology, and towards
higher temperatures which links QCD into the cosmological context.
Of particular recent interest is the lattice calculation of the topological
susceptibility in QCD, which provides the non-perturbative theoretical input
for axion cosmology.

\section{Equation of state at high temperatures}
\label{sec:ceos}

\begin{figure}{ht}
\begin{center}
\includegraphics[width=2.75in]{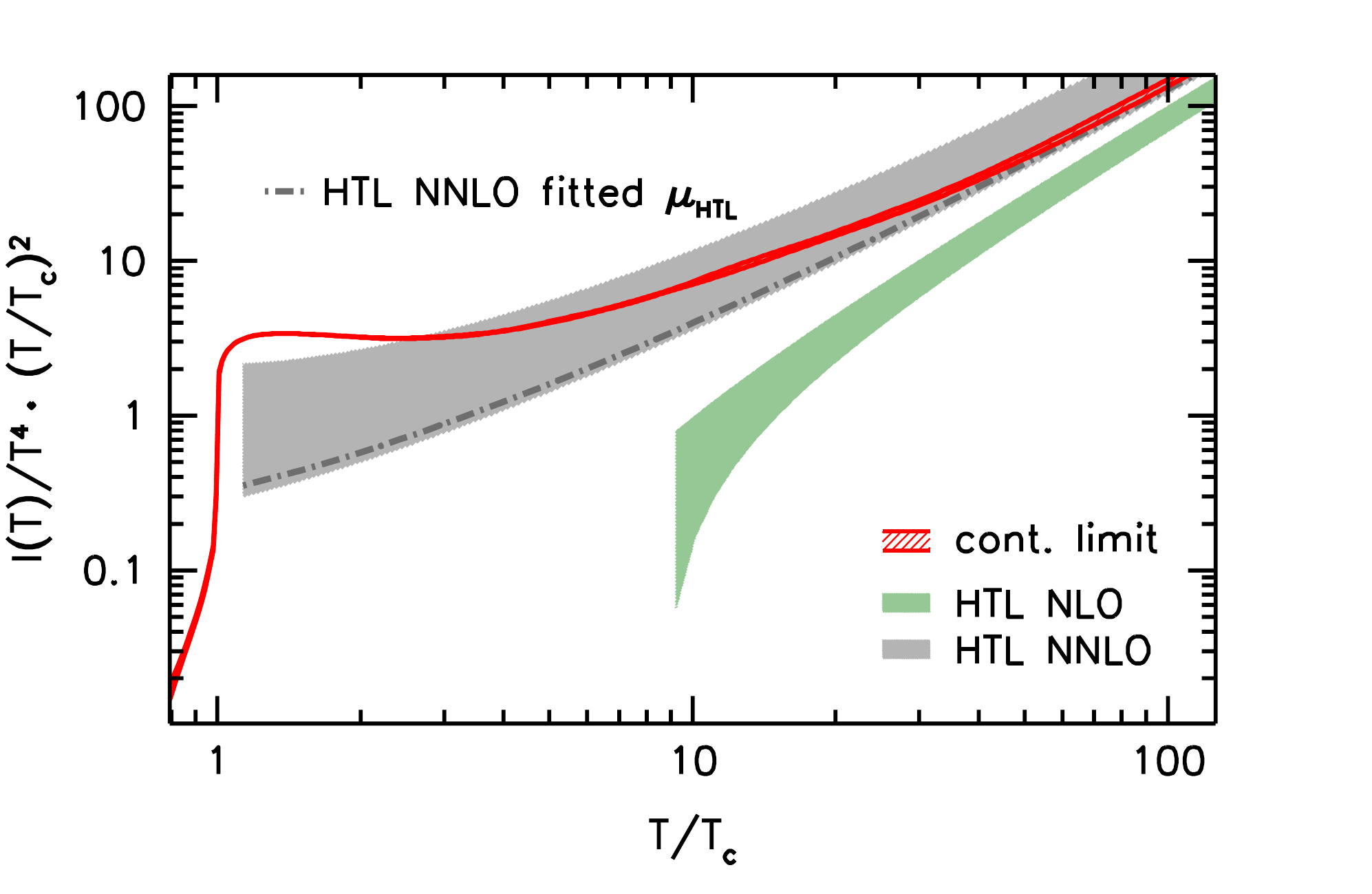}
%\hspace{0.25in}
\includegraphics[width=2.5in]{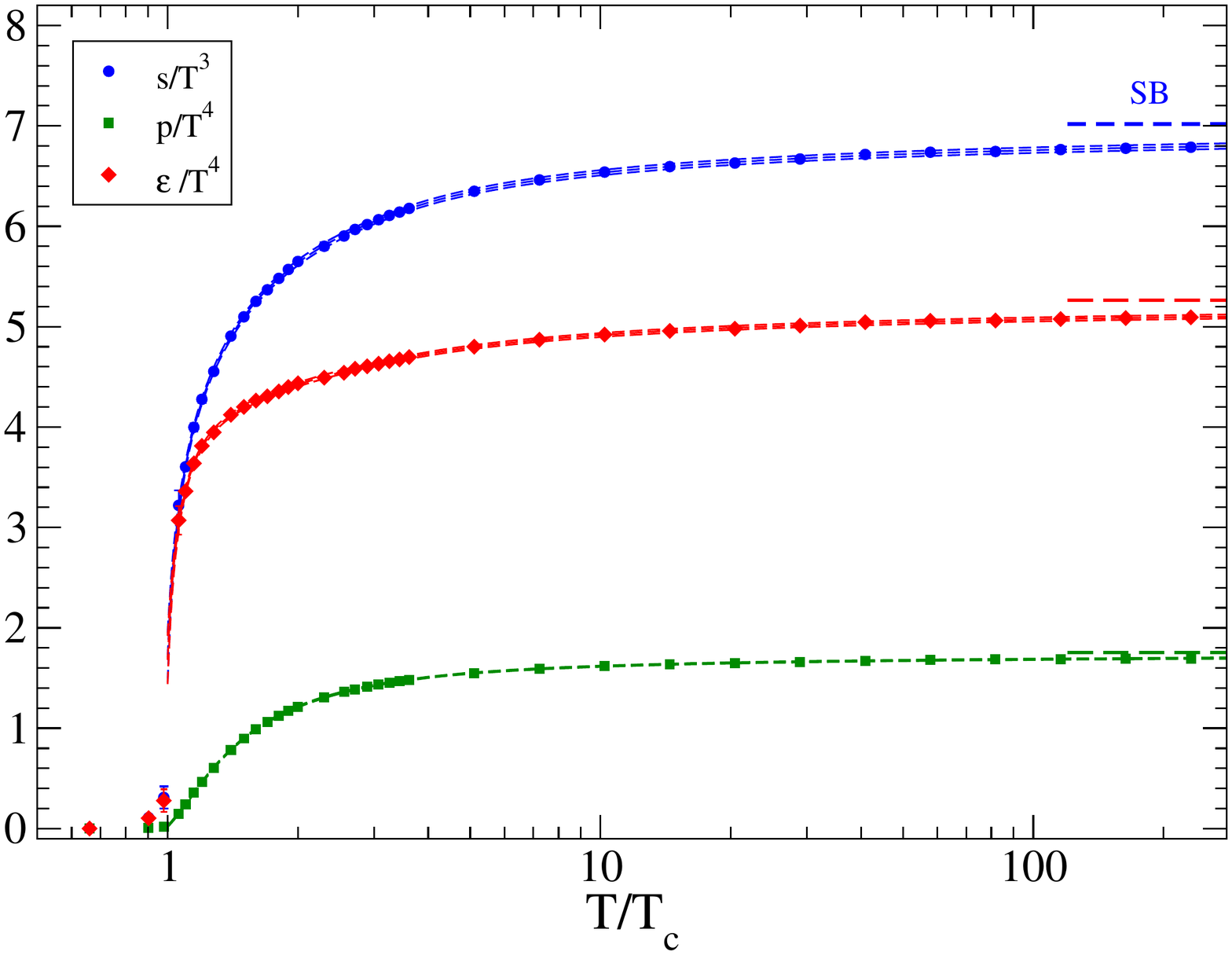}
\end{center}
\caption{\label{fig:qeos}
The equation of state in the quarkless theory in a broad temperature range.
The trace anomaly plot (left) was published by  the Wuppertal-Budapest group
from 2012 \cite{Borsanyi:2012ve}, this plot uses the non-standard normalization
$(\epsilon-3p)/T^2$.
The entropy density (blue), pressure (green)
and the energy density (red) on the right panel is taken from
Ref.~\cite{Giusti:2016iqr}, which is the latest publication on the subject.
}
\end{figure}

Even in the absence of quarks the study of high temperature lattice
fields becomes technically very difficult if the simulation volume is
not to shrink with the temperature. Typical simulation programs use
a given a lattice geometry of $N_\tau$ temporal slices in the Euclidean
time direction and $N_\sigma$ sites in each spatial directions. On
an isotropic lattice with spacing $a$, the temperature is $T=1/(N_\tau a)$,
and the box length is given by $L=aN_\sigma$. Thus, with fixed
geometry $L$ drops with $1/T$ as $L = N_\sigma/(N_\tau T)$.
In the pure SU(3) theory Refs.~\cite{Borsanyi:2012ve,Giusti:2016iqr}
used very large spatial lattices so that the simulations could be
carried beyond $10T_c$. While Ref.~\cite{Borsanyi:2012ve} used
the conventional technique \cite{Boyd:1996bx}  where the trace anomaly is
calculated from a pair of simulations at each temperature,
the approach of Ref.~\cite{Giusti:2016iqr} defines the ensemble in 
moving frame, allowing the extraction of the entropy from the off-diagonal
elements of the energy-momentum tensor \cite{Giusti:2012yj}. The latter
technique requires an elaborate renormalization procedure, which was worked out
in Ref.~\cite{Giusti:2015daa}. From the entropy function the other
observables are found through thermodynamic identities. We show the results in
Fig.~\ref{fig:qeos}.  Refs.~\cite{Borsanyi:2012ve,Giusti:2016iqr} agree for
high temperatures.  Below $3T_c$ disagreements are reported on the 4\% percent
level \cite{Giusti:2016wsf}.  Lacking systematic error budgets the significance
of these are difficult to assess.

A further interesting method was used in Ref.~\cite{Kitazawa:2016dsl},
which is based on the gradient flow. It was observed the signal-to-noise
ratio in the lattice energy-momentum operator can be significantly
enhanced by a filtering of the UV-modes, resulting in automatically
renormalized observables in the gauge sector
\cite{Luscher:2010iy,Luscher:2011bx}. A finite energy-momentum tensor was
obtained at positive flow times perturbatively \cite{Suzuki:2013gza},
these findings have already been generalized to full QCD \cite{Makino:2014taa}.
The first numerical results with Wilson quarks have been published in
Ref.~\cite{Taniguchi:2016ofw}. Though these directions are very promising,
an implementation with physical quarks masses is still missing.
So that we can discuss continuum extrapolated results with a physical
particle spectrum we turn to the staggered formulation.
 
\begin{figure}[t]
\begin{center}
\includegraphics[height=1.9in]{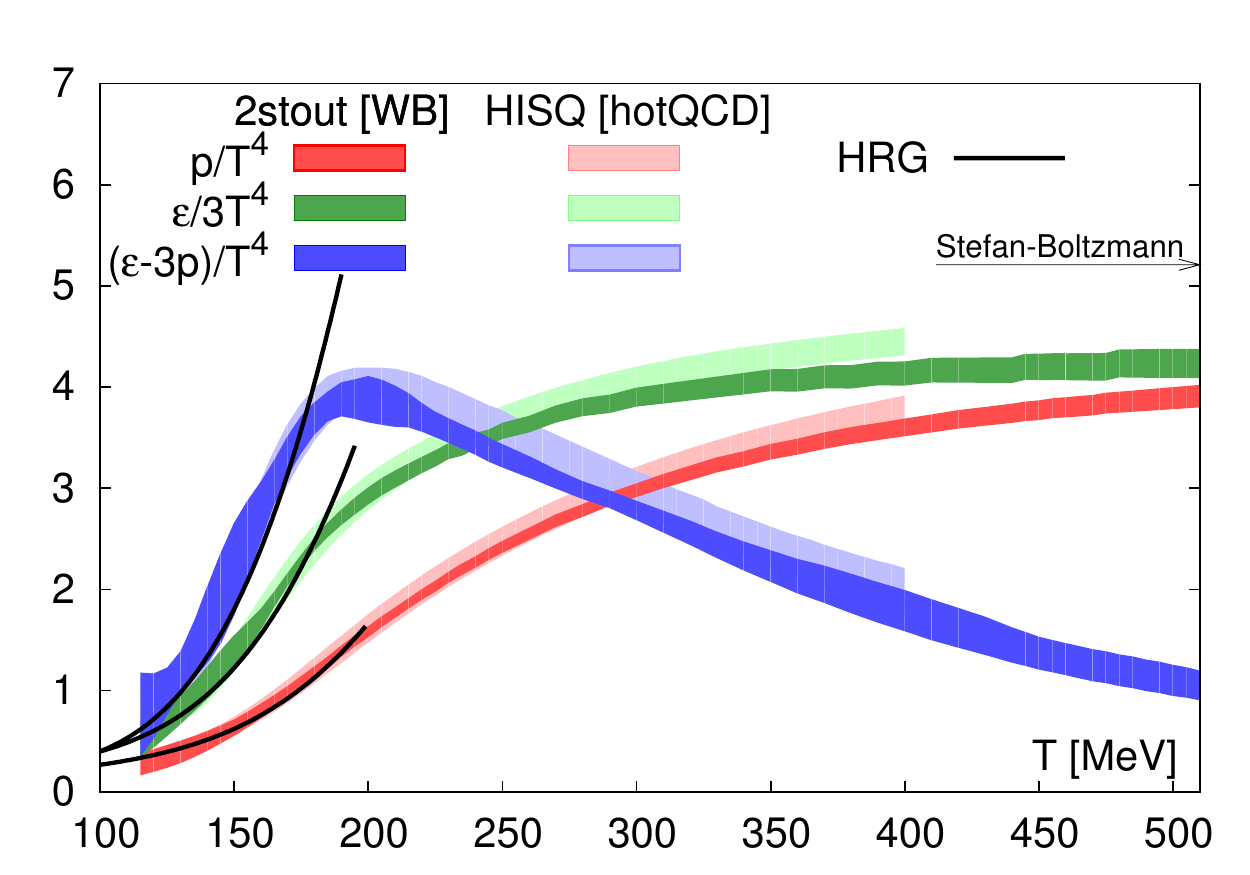}
\includegraphics[height=1.9in]{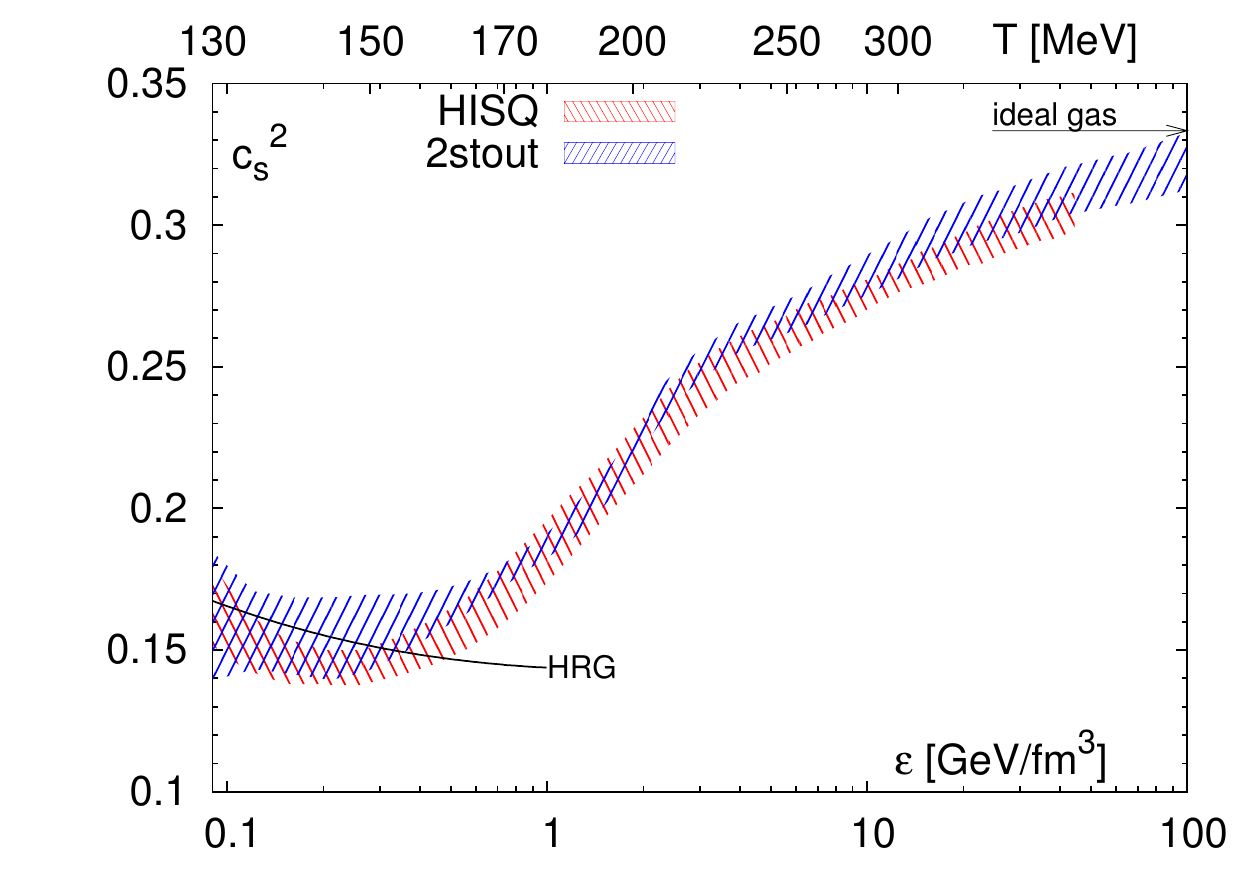}
\end{center}
\caption{\label{fig:pressure_etal}
{\em Left:} Continuum results for the pressure (red), energy density (green)
and trace anomaly (blue) as a function of temperature, as published by
the Wuppertal-Budapest group \cite{Borsanyi:2013bia} using the 2stout action
 and the HotQCD collaboration \cite{Bazavov:2014pvz} using the HISQ action.
{\em Right:}
The speed of sound ($c_s^2=dp/d\varepsilon$) plotted as a function of the
energy density. On the top axis we label the respective temperatures.
}
\end{figure}

The study of the equation of state in full QCD is by far more challenging.
In Ref.~\cite{Borsanyi:2010cj} a continuum estimate was given
using stout-staggered quarks with physical quark masses. Later almost the same
results were obtained after continuum extrapolation \cite{Borsanyi:2013bia}
or with a different type of staggered quarks \cite{Bazavov:2014pvz}. These
results are summarized in Fig.~\ref{fig:pressure_etal}.

The studies presented in Fig.~\ref{fig:pressure_etal} were based on the
simulations of two discretizations of the 2+1 flavor action. Thus
the effect of the charm was here neglected, although from about 300~MeV
temperature a significant contribution is expected based on
perturbative results \cite{Laine:2006cp}. Earlier partially quenched
works suggested an earlier onset of the charm 
\cite{Levkova:2009gq,Borsanyi:2010cj}, which was later explained as
a renormalization effect \cite{Borsanyi:2012vn}: in the equation of state
determination a key role is played by the $a(\beta)$ scale function
(with the inverse coupling $\beta=6/g_0^2$), which, too,
is significantly modified by the inclusion of the charm quark.

A satisfactory treatment of the charm degree of freedom, especially
if fine lattices are desired, required the introduction of a new lattice
action with dynamical charm. In Ref.~\cite{Borsanyi:2012vn} the already
used 2-stout staggered action was equipped with a charm quark. A year later
a new large scale staggered program was launched by the Wuppertal-Budapest
group, using four steps of stout smearing \cite{Morningstar:2003gk}, the
preliminary results for the equation of state were shown at the Quark Matter
conference in 2012 \cite{Ratti:2013uta}. The tuning procedure was refined
and the finite temperature ensembles were introduced in
Ref.~\cite{Bellwied:2015lba}.
The ETM collaboration also introduced the charm quark as the counterpart
of the strange quark, covering a pion mass range between 210 and 470~MeV.
They calculated the equation of state with 2+1+1
dynamical flavours for 370 MeV pions and three lattice spacings
\cite{Burger:2015xda}.
The MILC collaboration used the HISQ action with 2+1+1 dynamical flavors
and calculated the trace anomaly in Ref.~\cite{Bazavov:2013pra}. They
reported very small discretization effects with lattices as coarse as
$N_\tau=6,8$ and 10, the data were taken with a pion mass of 320 MeV.
This effort was, however, not pursued further towards the physical point.

In the rest of this section we present the result of the Wuppertal-Budapest
group, that published continuum extrapolated results with physical
quark masses in Ref.~\cite{Borsanyi:2016ksw}. For the equation of state
this paper used the same scheme as earlier in the quenched case 
\cite{Borsanyi:2012ve}. Each finite temperature ensemble is accompanied
with an other ensemble with the same bare parameters, but belonging to a
different temperature (different $N_\tau$). The quartic divergent contributions
are equal in the two simulations, the difference in the trace
anomaly ($I=\epsilon-3p$) contributions is $[I(T)-I(T/2)]/T^4$ assuming a
factor of two in $N_\tau$ between the two ensembles. These ensemble pairs come
supplementary to the $[I(T)-I(0)]/T^4$ pairs, which are conventionally
used, since at zero temperature the trace anomaly vanishes $I(0)=0$.
The spatial box size $L$ has been kept large even with increasing
temperatures, such that $LT_c\gtrsim 2$. E.g. at $T=880$~MeV 
we used the geometries $64^3\times 6$, $96^3\times 8$, $128^3\times10$
and $128^3\times 12$, and with the same bare parameters $I(T/2)$ was
calculated on $64^3\times 12$, $96^3\times 16$, $128^3\times 20$ and
$128^3\times24$ lattices. The ensembles up to 600~MeV temperature were listed in
Ref.~\cite{Bellwied:2015lba}.

\begin{figure}[ht]
\centerline{\includegraphics[width=2.5in]{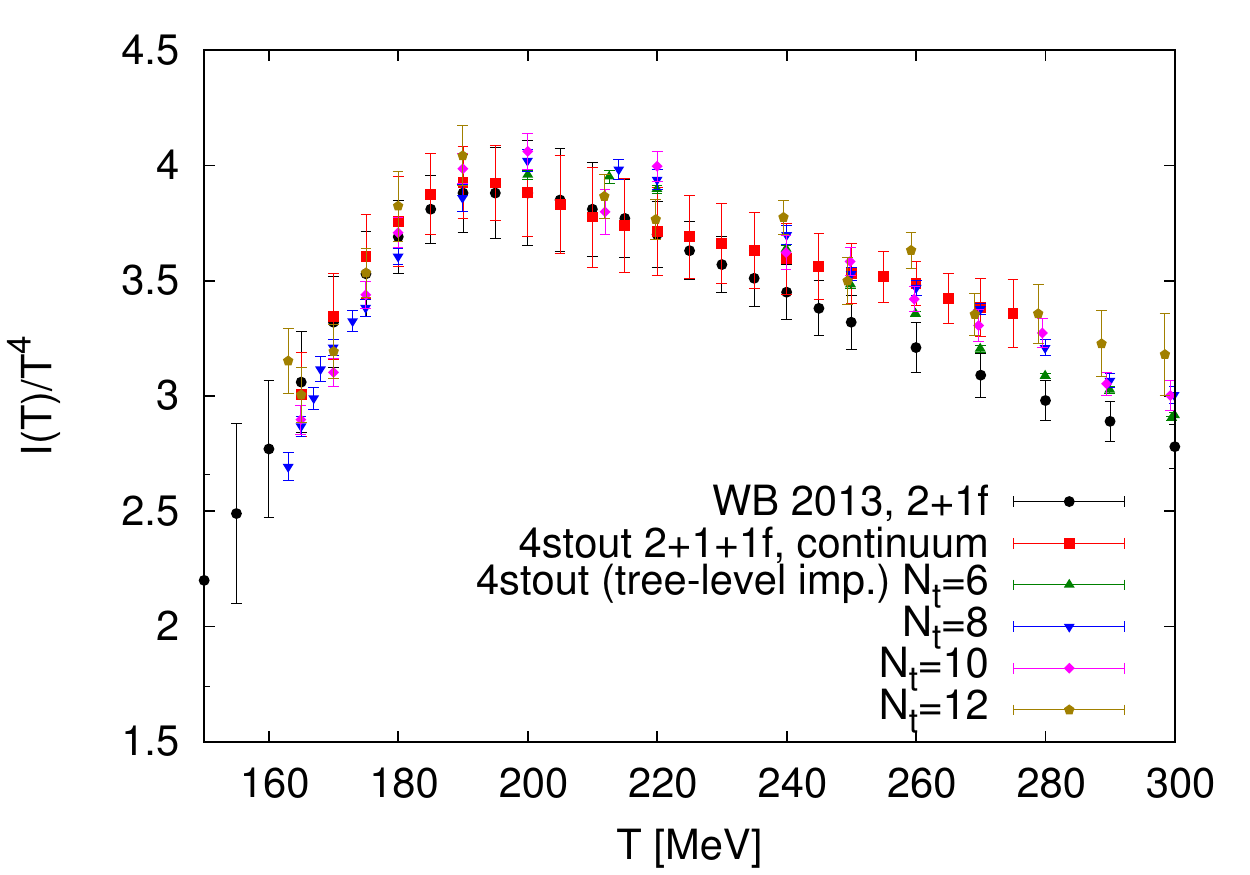}
\hspace{0.25in}
\includegraphics[width=2.5in]{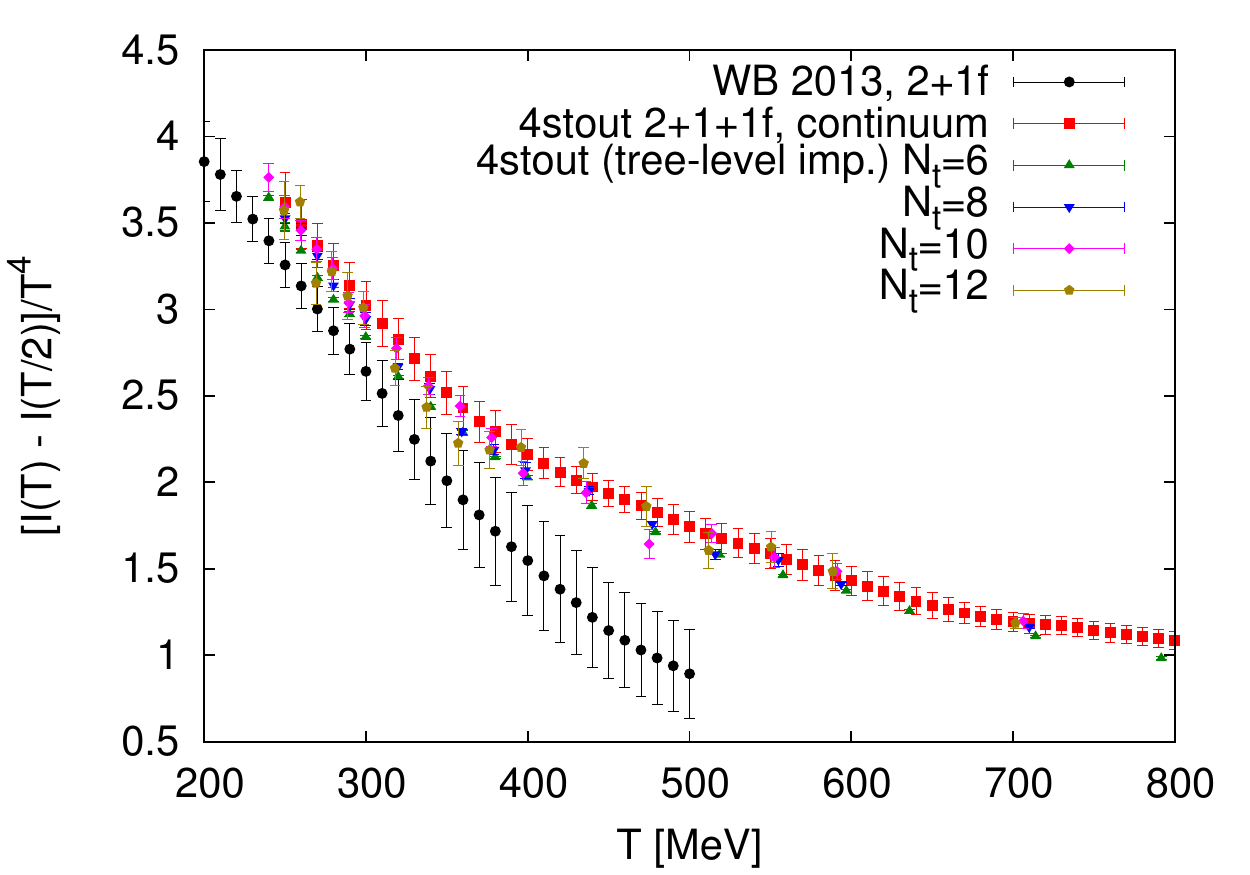}}
\caption{
\textit{Left panel:} the trace anomaly difference to zero temperature.
\textit{Right panel:} the trace anomaly difference to finite temperature.
The continuum extrapolation is based on $N_\tau=6,8,10$ and 12.
The full trace anomaly is calculated from Eq.~(\ref{eq:tracea}).
\label{fig:trah}
}
\end{figure}

The $[I(T)-I(0)]/T^4$ and the $[I(T)-I(T/2)]/T^4$ pairs are
shown before and after the continuum extrapolation in Fig.~\ref{fig:trah}.
Ref.~\cite{Borsanyi:2016ksw} then used the combined formula
\begin{equation}
\frac{I(T)}{T^4}=
\sum_{k=0}^{n-1}2^{-4k}\left.\frac{I\left(T/2^k\right)-I\left(T/2^{k+1}\right)}{\left(T/2^k\right)^4}
\right.
+
2^{-4n}\left.\frac{I\left(T/2^n\right)}{\left(T/2^n\right)^4}\right.\,,
\label{eq:tracea}
\end{equation}
to find the absolute $I(T)$ result where $n$ is the smallest non-negative
integer with $T/2^n <250~\mathrm{MeV}$. A comparison to the 2013 data set of
Ref.~\cite{Borsanyi:2013bia} shows that the effect of the charm is increasingly
relevant from about 300 MeV temperature.

It in natural to expect that towards 1 GeV temperature perturbation theory
becomes applicable. The low orders of the conventional perturbative series
fail to show convergence. Up to $\mathcal{O}(g^6\log g)$ the QCD 
pressure is known analytically \cite{Kajantie:2002wa}, the $\mathcal{O}(g^6)$
term can be found numerically through dimensional reduction to the three
dimensional effective theory \cite{Hietanen:2008tv}. Here the idea of
fitting the $\mathcal{O}(g^6)$ term directly to 4D data was also
discussed for the quark-less theory. This direct fit was later
recalculated as high temperature data became available \cite{Borsanyi:2012ve}.
A similar strategy was implemented now in the unquenched case. However,
because of the non-negligible charm quark mass the result could not be
fitted by the four-flavor perturbative result. Following a tree-level
version of the ratio in Ref.~\cite{Laine:2006cp} one writes

\begin{equation}
p^{(2+1+1)}(T)= \frac{\SB(3) + F_Q(m_c/T)}{\SB(4)}
{\left.p^{(2+1+1)}(T)\right|_{m_c=0}} \,.
\label{eq:charmcorr}
\end{equation}
Here $\SB(3)$ and $\SB(4)$ are the Stefan-Boltzmann limits of the three and four
flavour theories, respectively, and $F_Q(m_c)/T$ gives the free energy for one
free charm quark with $m_c=1.29~\mathrm{GeV}$. The leading perturbative
corrections to Eq.~(\ref{eq:charmcorr}) are calculated in
Ref.~\cite{Laine:2006cp}, these are well below the percent level near the
highest simulation point at about 1~GeV.

Thus we attempt to give a perturbative description to our simulation
results by fitting Eq.~(\ref{eq:charmcorr}) to the pressure result
at a single temperature $T_{\max}=1~\mathrm{GeV}$. Here we use
for $\left.p^{(2+1+1)}(T)\right|_{m_c=0}$
the known $\mathcal{O}(g^6\log g)$ formula plus one fit parameter ($q_c$)
controlling the $g^6$ contribution. The coupling is translated to temperature
using the four-loop running of the coupling with
$\Lambda_{\overline{MS}}=290$~MeV, the standard value for $n_f=4$
\cite{Agashe:2014kda}.

\begin{figure}[ht]
\centerline{
\includegraphics[width=2.5in]{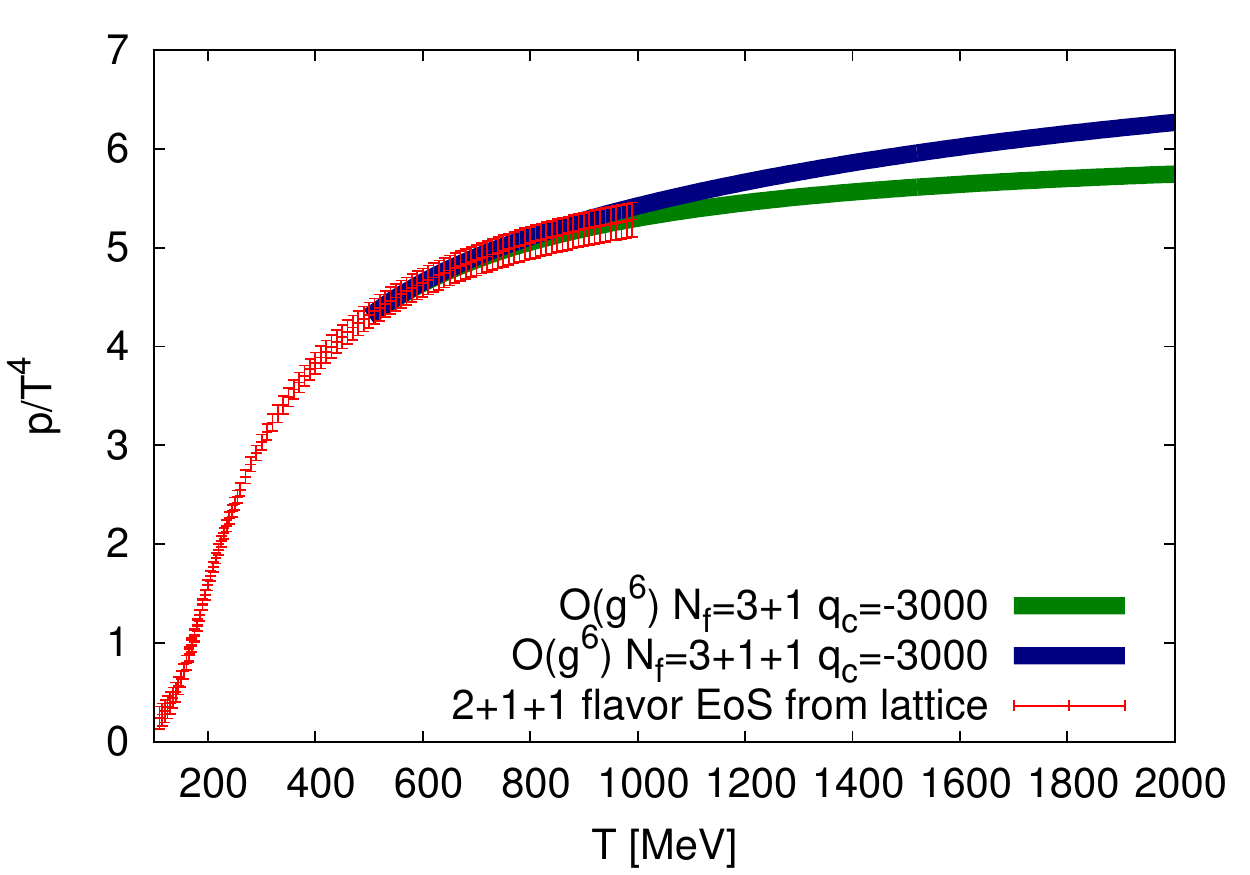}
\hspace{0.25in}
\includegraphics[width=2.5in]{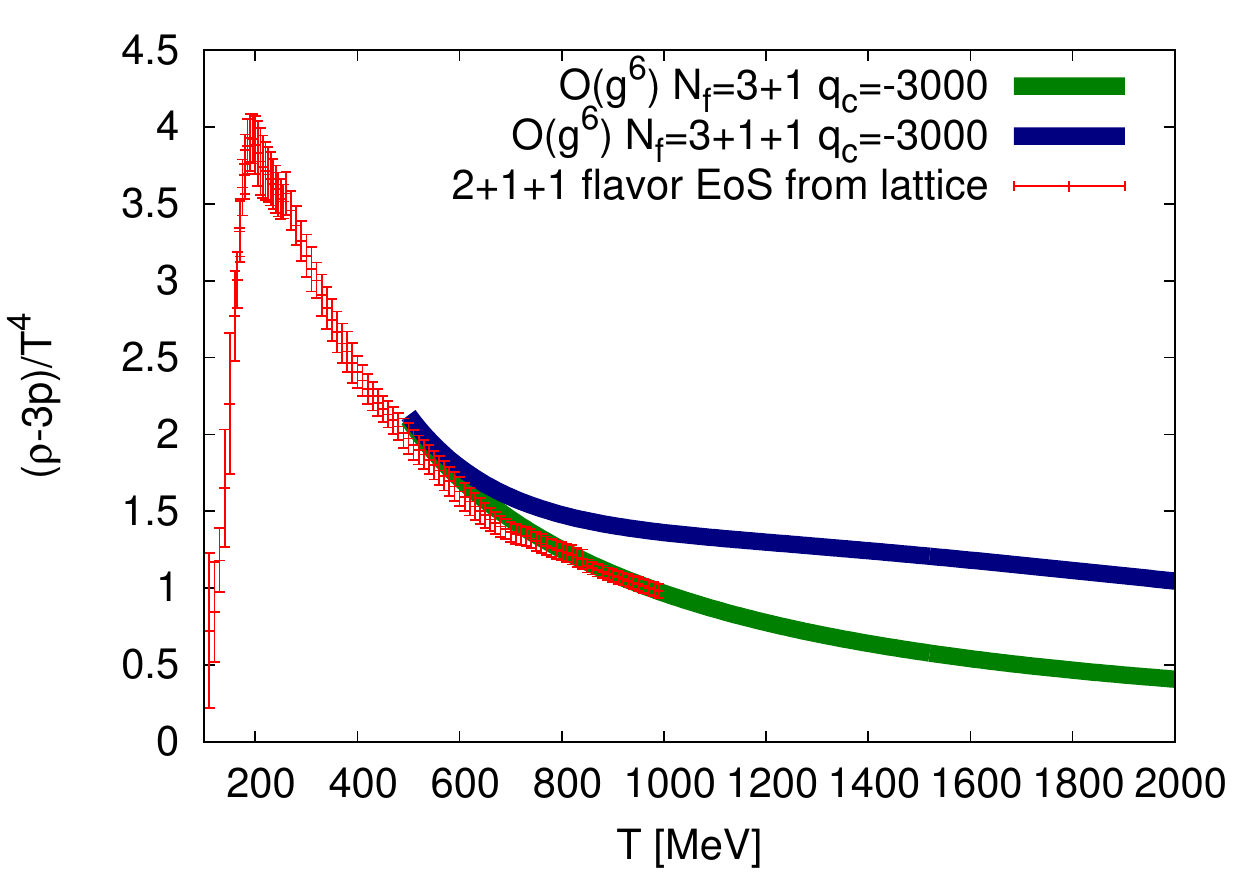}
}
\caption{\label{fig:withmass}
The continuum extrapolated lattice result is plotted in red 
for the 2+1+1 pressure (left) and trace anomaly (right) with 
physical quark masses. The single-parameter fit procedure results
in the green curves. When we include the contribution of the
bottom quark on the same footing we arrive at the result shown in blue.
}

\end{figure}

The non-trivial result is shown in Fig.~\ref{fig:withmass}.
Fixing the only fit parameter to $q_c=-3000$ (a parameter not far
from that in the pure Yang-Mills theory in Ref.~\cite{Borsanyi:2012ve}),
we have good agreement for both the pressure and the trace anomaly
starting already from approx 500~MeV. There is actually a $\pm10\%$
window in $q_c$ where a simultaneous agreement with the trace anomaly
and pressure is granted.  Since at high temperature 
the fitted formula is asymptotically equivalent to the known
perturbative result, one may expect that the agreement persists
for higher temperatures, too. A direct cross-check to lattice data
beyond 1 GeV is, however, technically demanding. Although
deep in the deconfined phase the simulation with dynamical quarks are
inexpensive, the autocorrelation times are increasingly difficult
to control. In addition, the systematics of the tuning of the bare
parameters (or in other words, the setting of the scale and the quark
masses) cannot be feasibly backed by zero-temperature simulations.

But once we demonstrated that simple parametrizations like
Eq.~(\ref{eq:charmcorr}) can be used to describe the quark mass effects
in the deeply deconfined phase we can attempt to go for higher
temperatures and add the bottom quark using the following ratio
\begin{equation}
p^{(2+1+1+1)}(T) = p^{(2+1+1)}(T) \frac{\SB(4) + F_Q(m_b/T)}{\SB(4)}\,
\label{eq:bottomcorr}
\end{equation}
where $m_b(m_b)=4.18$~GeV is the mass of the bottom quark \cite{Agashe:2014kda}.
Eq.~(\ref{eq:bottomcorr}) is expected to work beyond the bottom threshold.
Actually, the ratio of the four and five flavour perturbative results
are equal the respective ratio of their Stefan-Boltzmann limit to
about 0.3\% accuracy in the relevant temperature range. The blue line
in Fig.~\ref{fig:withmass} represent the l.h.s. of Eq.~(\ref{eq:bottomcorr}).
This is our best estimate to the QCD equation of state from 500~MeV temperature.
More details of the fitting procedure are given in Ref.~\cite{Borsanyi:2016ksw}.

In the same paper the pressure and energy density contributions
of photons, neutrinos, charged leptons are also added up. The
electroweak sector, including the $W^\pm$, $Z^0$ and Higgs
bosons as well as the top quark are taken from Ref.~\cite{Laine:2015kra}.
The full equation of state can then be used to close the Friedmann
equations and to determine the cooling rate of the Universe:
\begin{equation}
\frac{dT}{dt} = -\frac{\sqrt{24\pi \epsilon(T) } s(T) T}{M_{\rm pl} c_V(T) }
\label{eq:cooling}
\end{equation}

\section{Equation of state at finite density}
\label{sec:mueos}

An other extension of previous equation of state simulations is
towards finite densities. RHIC events freeze-out in a chemical
potential range between 20 and 400 MeV 
\cite{Andronic:2005yp,Andronic:2008gu}. Thus for the phenomenology
of the heavy ion collisions the exploration of the range up to
$\mu_B/T\lesssim 3$ is sufficient, for quark chemical potential
this means $\mu_{ud}/T\lesssim 1$. Conventional Monte Carlo
simulations at nonvanishing real chemical potentials are not possible,
since then the formerly positive definite fermion determinant may take
complex values, and thus the exponentialized action cannot be interpreted as a
probability weight. Moving the complex part into the observable
introduces heavily oscillating contributions, which is known as the sign
problem. On top of that, the most relevant configurations are probably
never sampled due to the ill-tuned action, this is the overlap problem.
An appropriate redefinition of the degrees of freedom have solved the
complex action problem in many other theories \cite{Gattringer:2016kco},
and recently much progress has been made to solve the overlap problem in gauge
theories using density of states methods \cite{Langfeld:2012ah}. Other
direct methods to solve the complex action methods include
Lefschetz thimbles \cite{DiRenzo:2015foa} and complex Langevin simulations
\cite{Aarts:2009hn,Fodor:2015doa}. Promising these new directions may be,
their applicability to full QCD has not yet been proven.

For small chemical potentials, however, the leading coefficients of a Taylor
expansion may already provide a satisfactory description of this range of the
QCD phase diagram. Calculations to $\mu^6$ order have already been
available for a decade \cite{Allton:2005gk}, the leading $\mu^2$ order
was explored two decades ago \cite{Gottlieb:1987ac}. These works were
neither with physical parameters, nor continuum extrapolated, though. It
turned out that the technical difficulty of getting the Taylor coefficients of
the free energy greatly increased as one moved on to lighter pions and
to finer lattices. The reduction of the lattice spacing and the introduction of
fat links (HISQ or stout) have finally allowed the calculation of the
continuum limits of the leading $\mu$ derivatives of the effective action
\cite{Borsanyi:2011sw,Bazavov:2012jq} as well as the leading Taylor
coefficient for the equation of state \cite{Borsanyi:2012cr}.

It has been a great challenge to achieve an acceptable signal for the
higher coefficients in realistic lattice simulations. For this the
fourth and sixth order fluctuations of conserved charges had to be
calculated. The determination of these fluctuations have been also motivated by
the experimental availability of ratios of cumulants
\cite{Adamczyk:2013dal,Adamczyk:2014fia}. The first applications of lattice QCD
have provided estimates for the freeze-out parameters
\cite{Bazavov:2012vg,Borsanyi:2013hza,Borsanyi:2014ewa}.  The extrapolation of
the fluctuations to finite chemical potential may allow the extraction of
the curvature of the freeze-out line from fluctuation data
\cite{Bazavov:2015zja,Bellwied:2016cpq,Karsch:2016yzt}.

\begin{figure}[ht]
\centerline{
\includegraphics[width=2.5in]{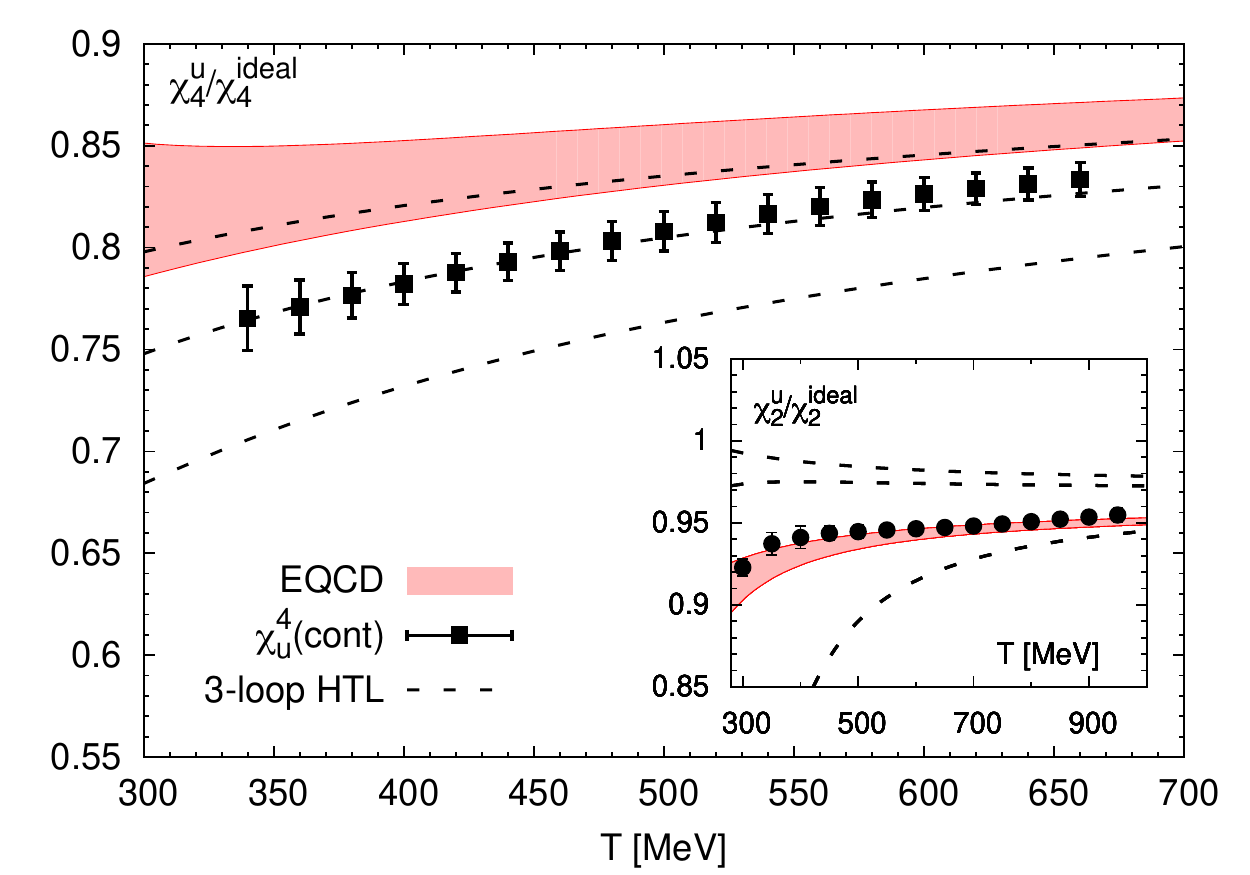}
\hspace{0.25in}
\includegraphics[width=2.5in]{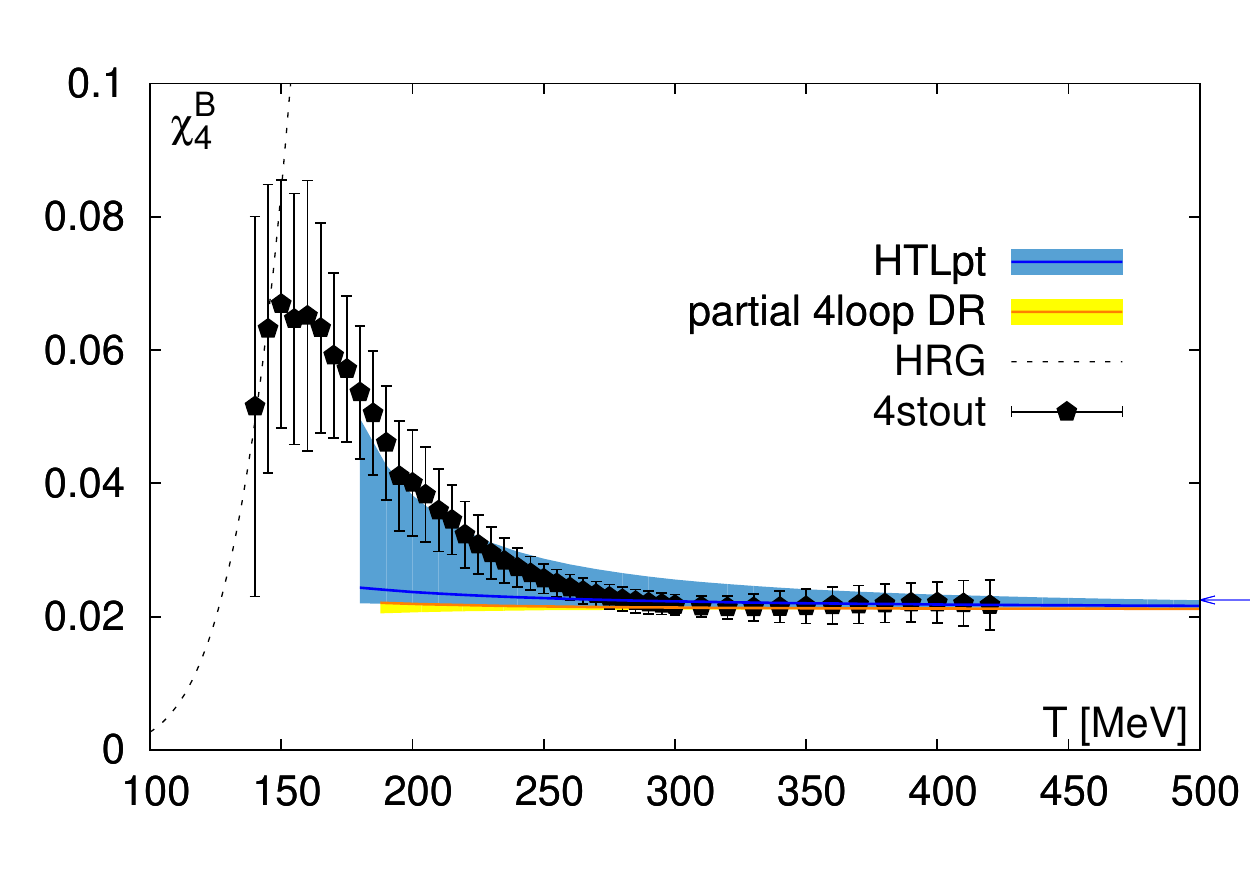}
}
\caption{\label{fig:highsusc}
The continuum extrapolated lattice result for the fourth
order fluctuation of the up quark number (\textit{left}) \cite{Ding:2015fca}
and the analogous fluctuation for the baryon number (\textit{right})
\cite{Bellwied:2015lba}.
In both plots the data are confronted to the expectations based
on HTL perturbation theory \cite{Haque:2014rua} and dimensional reduction
\cite{Mogliacci:2013mca}.
}
\end{figure}

The fourth-order derivatives in the continuum limit have been published
in Ref.~\cite{Bellwied:2015lba}. In Fig.~\ref{fig:highsusc} we show $\chi^U_4$,
the fourth derivative of the free energy density with respect to the up quark
chemical potential (the kurtosis of the up quark number) from
Ref.~\cite{Ding:2015fca} and the analogous quantity for the baryon number
$\chi^B_4$ from Ref.~\cite{Bellwied:2015lba}. In both cases the data
are compared to the predictions based on hard thermal loop perturbation
theory \cite{Haque:2014rua} and to those from dimensional reduction
\cite{Mogliacci:2013mca}.

\begin{figure}[ht]
\centerline{
\includegraphics[width=\textwidth]{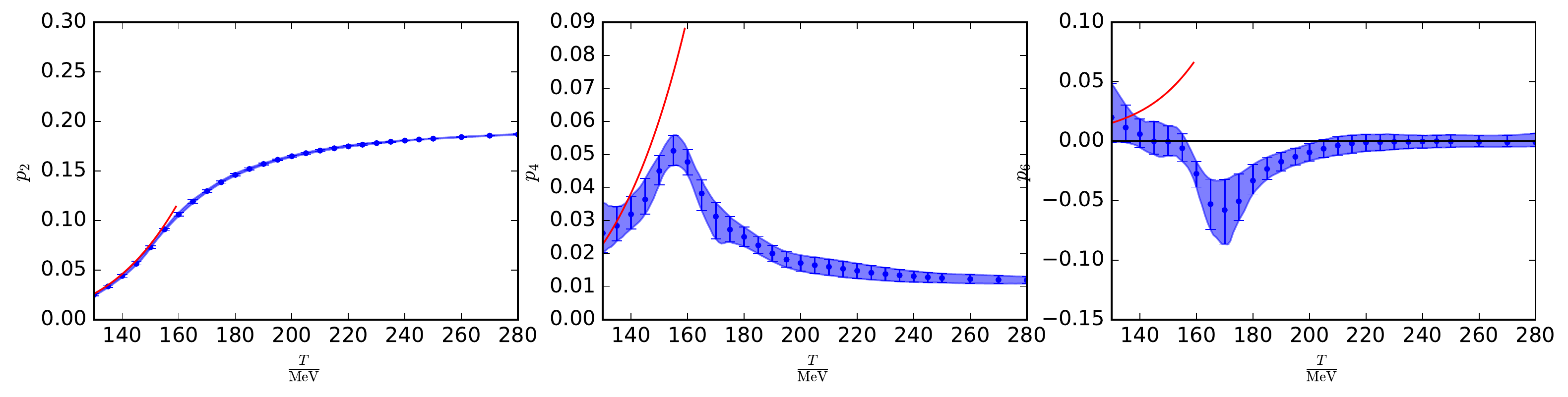}
}
\caption{\label{fig:coefficients}
Taylor coefficients of the QCD pressure in the transition region
calculated from imaginary chemical potential data
by the Wuppertal-Budapest group \cite{Gunther:2016vcp}.
Similar results were presented at the Confinement12
conference by C. Schmidt using BNL-Bielefeld-CCNU data at $\mu_B=0$.
}
\end{figure}

From the error bars in Fig.~\ref{fig:highsusc} one can quickly notice that
the calculation costs of the fluctuations (and, hence, for the Taylor
coefficients, too) rapidly increases as the temperature is lowered towards
and below $T_c$. To meet the challenge, the BNL-Bielefeld-CCNU group invested
in a large pool ($\sim 10^5$) of configurations for each temperature,
although mostly at the resolution $N_\tau=8$. The Wuppertal-Budapest group
collected statistics at $N_\tau=10,12$ and 16, mainly using imaginary
chemical potentials in the simulation parameters. There is no complex action
problem hindering simulations with imaginary values of the chemical potentials.
As it was recently emphasized in Ref.~\cite{DElia:2016jqh} knowing the
$\mu_B$-dependence of low order $\mu_B$ derivatives of the free energy
allows the calculation of higher derivatives. This principle was exploited
by the Wuppertal-Budapest group in Ref.~\cite{Gunther:2016vcp}. 
Both in the Wuppertal-Budapest simulation program
\cite{Bellwied:2015rza,Bellwied:2016cpq} as well as in
the BNL-Bielefeld-CCNU effort \cite{Hegde:2014sta,Karsch:2016yzt} care
has been taken to correctly take into account the mixing of the strange
and baryon chemical potentials. To reproduce the experimental setting,
the expectation value of the strangeness has to be zero, and the electric
charge is set to 0.4 times the baryon number to reflect the flavor
imbalance of the projectile ions. In Ref.~\cite{Gunther:2016vcp} it
meant to match the imaginary strangeness chemical potential to the imaginary
baryon chemical potential, thus setting a trajectory in the imaginary
$\mu_B-\mu_S$ plane. The Taylor coefficients then correspond to the
total derivatives along this trajectory. The results are plotted in
Fig.~\ref{fig:coefficients}.

\section{Topological susceptibility at high temperatures}
\label{sec:topsusc}

Perhaps the most exciting topic of the past year was the topological
susceptibility of QCD at finite temperature. It is defined as the
second derivative of the free energy with respect to the $\theta$
parameter, which is introduced in the Euclidean theory as
\begin{equation}
\mathcal{L}_{E,\theta} = 
-i\theta \frac{g^2}{64\pi^2} \epsilon_{\mu\nu\rho\sigma}
F^a_{\mu\nu}(x) F^a_{\rho\sigma}(x)\,.
\end{equation}
A nonvanishing $\theta$ parameter would introduce an electric
dipole moment for the neutron \cite{Baluni:1978rf,Crewther:1979pi},
but this is constrained experimentally \cite{Baker:2006ts}, such that
$|\theta|<10^{-9}$. Given the CP-violation coming from the CKM-matrix
a fine-tuning is necessary to maintain a near-zero $\theta$ term,
unless at least one quark has zero mass, in which case the $\theta$
term can be removed by chiral rotation. However, there is no
quark with zero mass \cite{Durr:2010vn}. 

To solve this fine-tuning problem, the strong CP problem, Peccei and Quinn
have proposed to introduce dynamics to $\theta$ \cite{Peccei:1977hh}. 
Assuming a spontaneously broken global U(1) symmetry for a new dynamical 
field, the angular degree of freedom will the play of $\theta$.
This pseudo-Goldstone boson acquires a small mass ($m_a$) by the QCD chiral
anomaly. This mass
is then proportional to the second derivative of the effective potential:
\begin{equation}
m_a^2(T) f_a^2 = \frac{\partial^2 V_{\rm eff}}{\partial \theta_a^2}=\chi_t(T)
\end{equation}
where $f_a$ is the U(1) symmetry breaking scale, yet to be determined. This
angular degree of freedom is the axion field (for a brief review see
Ref.~\cite{Kim:2009xp}). The topological susceptibility $\chi_t$ at zero
temperature is well predicted by chiral perturbation theory
\cite{Leutwyler:1992yt}, and can be calculated in quenched
\cite{DelDebbio:2004ns,Durr:2006ky} and in the unquenched theory
\cite{Bernard:2003gq,Bruno:2014ova}.  Early lattice studies have shown that the
topological susceptibility sharply drops at the transition temperature
\cite{Alles:2000cg,Gattringer:2002mr}. The strong temperature dependence
can be explained by the dilute instanton gas approximation (DIGA)
\cite{Gross:1980br,Pisarski:1980md}. In a simplified picture the
action of a single instanton $2\pi/\alpha_s$ contributes as
\begin{equation}
\chi_1(T)\sim T^4 e^{-2\pi/\alpha_s(T)}
\end{equation}
to the topological susceptibility. From the one-loop running of the coupling we
have 
\begin{equation}
e^{-2\pi/\alpha_s}\sim T^{-11+2/3 N_f}\,.
\end{equation}
Actually, in the presence of fermions the configurations with non-trivial
topology are suppressed by a factor of the light quark mass, which,
at high temperature, is made dimensionless as $\sim m/T$. The contributions 
together give a power law:
\begin{equation}
\chi(T)\sim 1/T^{4-11+2/3 N_f-N_f}
\end{equation}
For a full calculation see Ref.~\cite{Ringwald:1999ze}.

\begin{figure}
\centerline{
\includegraphics[width=2.5in]{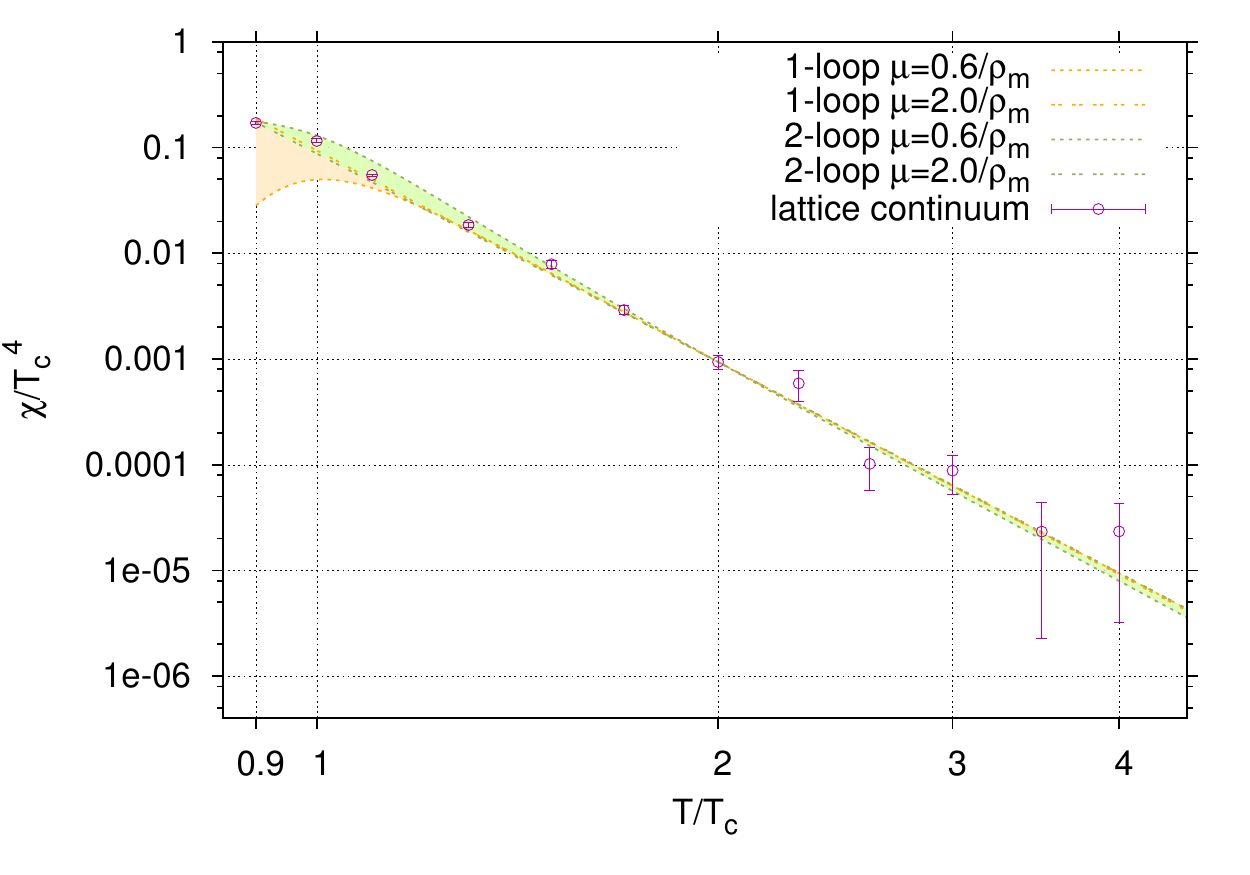}
\includegraphics[width=2.5in]{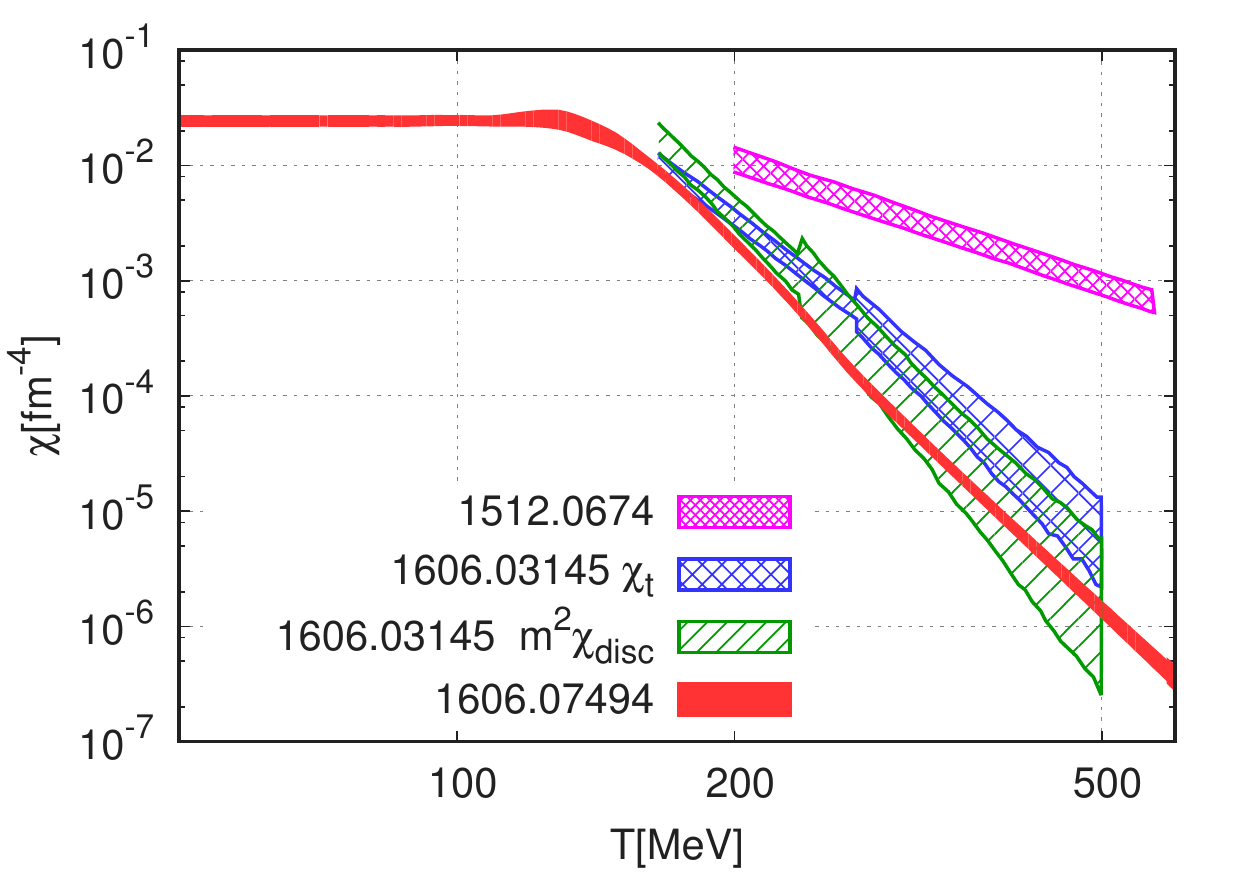}
}
\caption{\label{fig:topsusc}
Topological susceptibility as a function of temperature.
Left: Quenched results are compared to the DIGA prediction.
Here a pre-factor had to be introduced for DIGA to match
the lattice data \cite{Borsanyi:2015cka}. 
Right: A compilation of recent results for full QCD.
Magenta band: Bonati et al \cite{Bonati:2015vqz} using
2-stout staggered fermions and the gluonic definition of the
topological charge. Blue and green bands by Petreczky et al
\cite{Petreczky:2016vrs} come from the gluonic definition and the disconnected
chiral
condensate, respectively, using HISQ fermions.
The red band by the Wuppertal-Budapest group \cite{Borsanyi:2016ksw}
comes from a combination of simulations with staggered and overlap
quarks.
}
\end{figure}

To demonstrate the power law behaviour on lattice was a challenge
even in the quenched theory \cite{Berkowitz:2015aua,Kitano:2015fla,
Borsanyi:2015cka}. The result is shown on the left panel of
Fig.~\ref{fig:topsusc}. The power law $\chi\sim T^{-7}$ behaviour means that as
the temperature is increased in a constant volume the weight
of the configurations with non-zero topology shrinks with the same power,
and the length of the Monte Carlo simulations must be extended accordingly.
Moreover, an algorithmic difficulty, the topological freezing, preventing the
crossing between topological sectors, worsens the problem.
Suggestions include artificial decreasing of the weight of the trivial
sector \cite{Kitano:2015fla} and the use of the topological charge
density correlator \cite{Bautista:2015yza}.

In Refs.~\cite{Frison:2016vuc,Borsanyi:2016ksw} the following new
strategy was introduced. The partition sum is actually a sum over
the contributions from different topological sectors with charge $Q$:
\begin{equation}
Z = \int DU e^{-S}
 = \sum_Q \int \left.DU\right|_Q e^{-S}
 = \sum_Q Z_Q
\end{equation}
In a given volume at high temperature $Z_0 \gg Z_1 \gg Z_2 \dots$, and
through $\mathcal{P}$-symmetry: $Z_1=Z_{-1}$, $Z_2=Z_{-2},\dots$
If we keep the simulation aspect ratio $R=LT$ fixed, the susceptibility
can be written as
\begin{equation}
R^3\frac{\chi_t}{T^4} = \langle Q^2 \rangle = \frac{\sum_Q Q^2 Z_Q}{\sum_Q Z_Q}
=\frac{ 0 \cdot Z_0  + 1 \cdot Z_1 + 1 \cdot Z_{-1} + \mathrm{small}}{Z_0+\mathrm{small}}=\frac{2Z_1}{Z_0}
\end{equation}
Thus the temperature dependence of the topological susceptibility can
be deduced from the trace anomaly difference between the sectors.
\begin{equation}
\frac{\partial \log \chi_t}{\partial \log T}  =  4 +
\frac{\partial \log Z_1}{\partial \log T}  -
\frac{\partial \log Z_0}{\partial \log T}
\end{equation}
This way the  exponent of the power law behaviour
is obtained directly. In full QCD this amounts to:
\begin{equation}
{\chi_t ~\sim T^{-b}},\quad \mathrm{with}\,\,
b=-4-\frac{d\beta}{dT}\langle S_g\rangle_{1-0}
-\sum_f \frac{d m_f}{dT} m_f \langle \bar\psi\psi_f \rangle_{1-0}
\end{equation}
Here we used the notation $\langle \cdot \rangle_{1-0}$ for the difference
between the expectation values in the $Q=1$ and $Q=0$ sectors.

\begin{figure}
\centerline{
\includegraphics[width=2.5in]{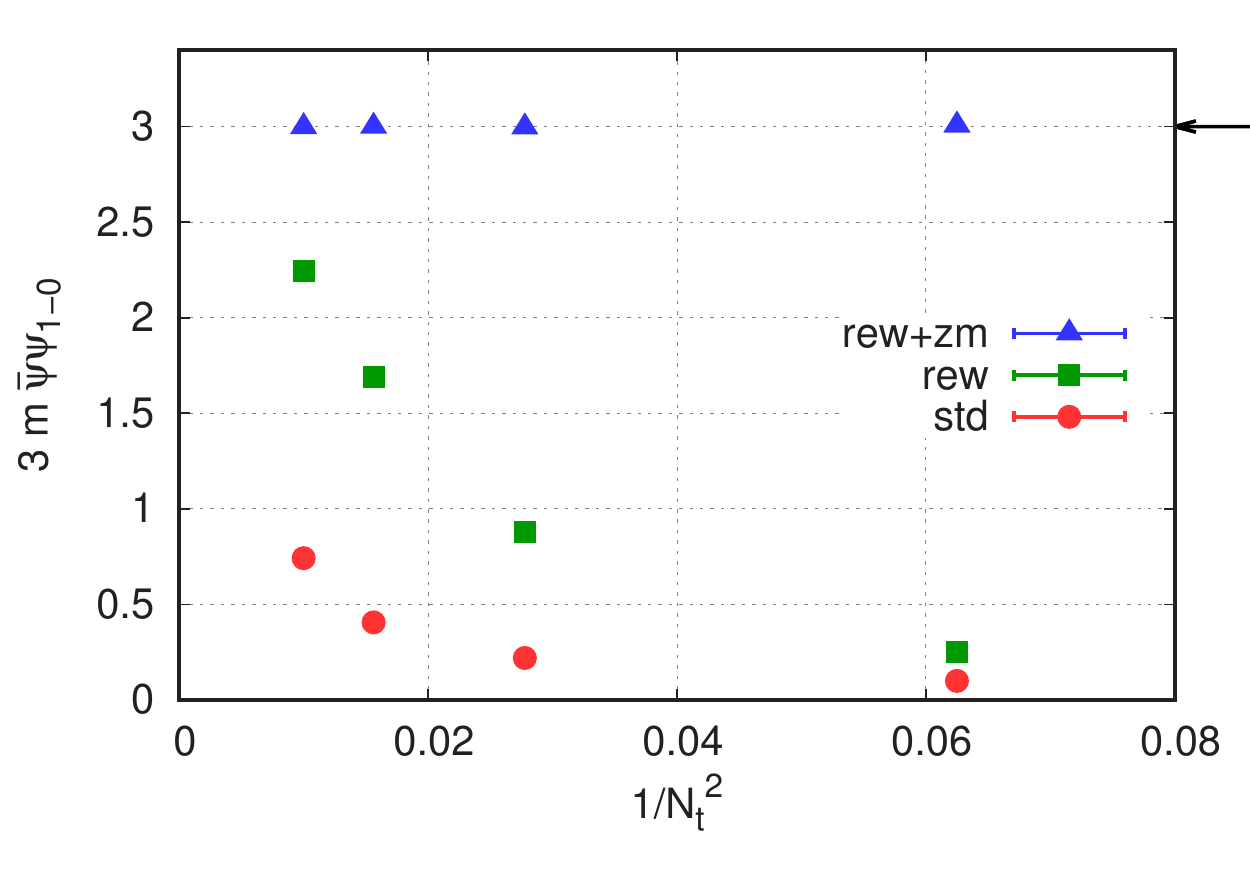}
\includegraphics[width=2.5in]{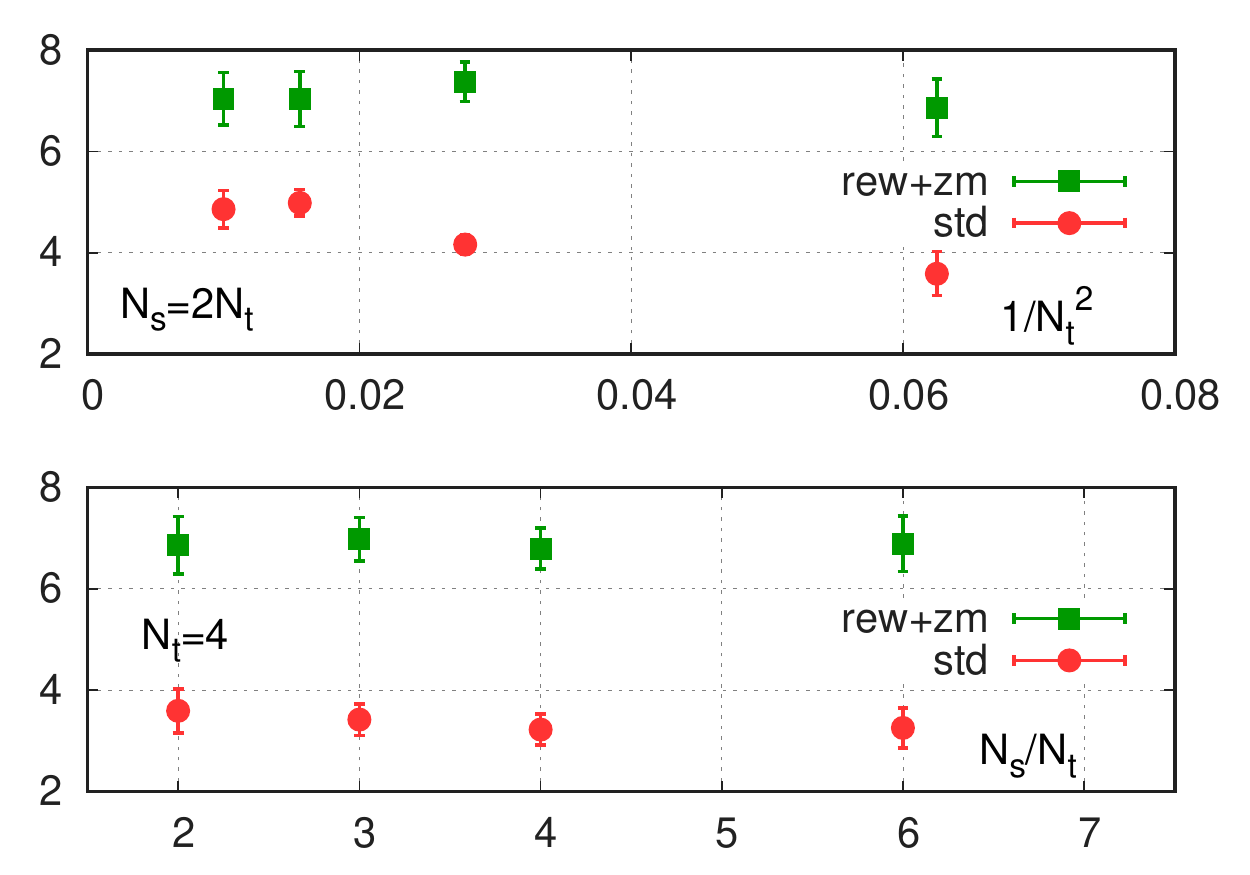}
}
\caption{\label{fig:continuum}
Left: the quark condensate difference between the $Q=1$ and $Q=0$ sub-ensembles
for various lattice spacings with the 4-stout staggered action at 750 MeV
temperature using 3+1 flavors. 
The red dots show the result in an unimproved ensemble. We get the green
squares after reweighting and the blue triangles after reweighting and the
consistent redefinition of the zero-mode contribution to the lattice
$\bar\psi\psi$ operator. The latter version shows a spectacular improvement
in the continuum scaling. The result is very close to the high temperature limit, 3, marked by the arrow.
Right: the exponent $b$ in $\chi_t\sim T^{-b}$ is calculated using
and not using reweighting. On the lower panel we show the sensitivity 
to the choice of volume.
All plots were published in Ref.~\cite{Borsanyi:2016ksw}.
}
\end{figure}

However, the generalization to full QCD is not as straightforward.
Discretization errors are severe, they can only be slightly reduced
by advantageous normalization, e.g. to the zero temperature $\chi_t(T=0)$ at the
same lattice spacing \cite{Bonati:2015vqz,Borsanyi:2016ksw}.
One very important lattice artefact in the staggered formalism
is the enhancement of the near-zero eigenvalues in the Dirac operator.
The number of the zero modes is related to the topological charge
though the index theorem. Even a large discretization error on a few
eigenvalues in the scarce non-trivial configurations has probably
small effect on other bulk quantities, for the topological susceptibility,
however, the very weight of these configurations matter. To correct the weight,
a reweighting strategy was suggested in Ref.~\cite{Borsanyi:2016ksw}.
Depending on $Q$, the staggered eigenvalues of the 
appropriate number of low modes are replaced by the would-be eigenvalues,
given by the mass. Since only the weight of the configurations are affected,
it is irrelevant whether the would-be zero mode or a mode with a very similar
eigenvalue was picked. Note that the fermionic observables, e.g. 
$\langle \bar\psi\psi_f \rangle_{1-0}$ also need to be redefined according
the modified action \cite{Borsanyi:2016ksw}. In Fig.~\ref{fig:continuum}
we show examples for the improved continuum scaling as well as a 
brief finite volume study.

The lattice results on the topological susceptibility are summarized
in Fig.~\ref{fig:topsusc}. The left panel shows the continuum extrapolated
quenched result together with the DIGA prediction. Although DIGA
did predict the exponent correctly, an order of magnitude mismatch
is found in pre-factor \cite{Borsanyi:2015cka}. The full QCD result by three
collaborations \cite{Bonati:2015vqz,Petreczky:2016vrs,Borsanyi:2016ksw} 
is shown in the right panel of Fig.~\ref{fig:topsusc}.

Having determined the QCD topological susceptibility over a broad
range of temperatures the only missing parameter for the temperature-dependent
axion mass is the $f_a$ constant. For this several cosmological constraints
exist \cite{Raffelt:2006cw}. To derive these model dependent assumptions
for the axion couplings have to be made, e.g. using the DFSZ \cite{Dine:1981rt}
or the KSVZ \cite{Kim:1979if,Shifman:1979if} model. The strongest
lower bound is coming from the SN~1987A supernova event, suggesting $f_a\gtrsim
10^{-9}~\mathrm{GeV}$ \cite{Raffelt:2006cw}. An upper bound is set by the
known abundance of the cold dark matter. Assuming a random alignment for
the axion field in the expanding Early Universe the axion zero mode obeys
the damped oscillator equation
\begin{equation}
\ddot \theta + 3H \dot \theta + m_a^2(T)\sin\theta =0\,.
\end{equation}
It is initially completely dominated by the Hubble friction term. Only
as the temperature drops and, thus, the topological susceptibility $\sim m_a^2$
rises such that  $m_a\approx 3H$ will the oscillatory dynamics start.
Following the cooling rate of the universe from Eq.~(\ref{eq:cooling})
and assuming an adiabatic evolution of the oscillations in the time-dependent
axion potential the evolution of the axion density can be predicted.
The energy density today mostly depends on the entropy ratio between today
and the starting point of the axion oscillations. Through this mechanism, the
misalignment mechanism, the Universe is populated with axions, their density
is strictly bound from above by the dark matter abundance \cite{Wantz:2009it}.

Axions, if they exist, might provide only a small fraction of the dark matter,
especially since the axionic string formation may also be non-negligible.
Using the newly calculated topological susceptibility one can now calculate
the cosmological axion bound. An axion mass of
$m_a=28(2)\mu\mathrm{eV}$ represents a lower bound, in that case the
axions from the misalignment mechanism could account for 100~\% of dark matter,
or, if the axion mass is $m_a=50(4)\mu\mathrm{eV}$ 50~\% percent is accounted
for.  If one assumes that axions from the misalignment mechanism 
contribute between 1 and 50~\% to dark matter, this range translates to a mass
range of $m_a=50 - 1500 \mu\mathrm{eV}$ \cite{Borsanyi:2016ksw}.

\section{Summary}
\label{sec:summary}
In the past years the finite temperature 
lattice calculations have been extended to several
new observables. Here we focused on the equation of state, and fluctuations
and the topological susceptibility. The latter direction was the most
demanding from the technical point of view, where severe discretization
errors, an extremely low signal to noise ratio and long autocorrelation times
had to be defeated. We hope that progress reported here will continue and
soon the higher moments will be determined both in chemical potential
and in the $\theta$ dependence.

\bibliographystyle{woc.bst}
\bibliography{thermo}{}
\end{document}